\documentstyle[aps]{revtex}
\begin{document}
\bibliographystyle{prsty}
\baselineskip=5.5mm
\parindent=7mm
\begin{center}
{\large {\bf \sc{$B-\pi$ weak form factor with chiral current
in the light-cone sum rules }}} \\[2mm]
Zhi-Gang Wang$^{1,2}$ \footnote{E-mail,wangzg@mail.ihep.ac.cn}, Ming-Zhen Zhou$^{2}$, Tao Huang$^{1,2}$    \\
$^{1}$ CCAST (World Laboratory), P.O.Box 8730, Beijing 100080,
P.R.China \\
$^{2}$ Institute of High Energy Physics, P.O.Box 918 ,Beijing 100039,P. R. China\footnote{Mailing address } \\
\end{center}

\begin{abstract}
In this article, we calculate the
$B\to \pi$ transition form factor $f^{+}_{B\pi}(q^2)$ by
including  perturbative $O(\alpha_s)$ corrections to the
twist-2 terms with chiral current in the light-cone QCD sum rule approach.
The corrections to the product $f_Bf^{+}_{B\pi}(q^2)$
in the leading twist approximation is found to be about $30\%$,
while a similar magnitude corresponding to $O(\alpha_s)$ corrections
 for $f_B(q^2)$ in the two-point sum rule cancel them and result in
  small net corrections for $f^{+}_{B\pi}(q^2)$. Our results confirm the
  observations made in previous light-cone QCD sum rule studies.
\end{abstract}



\section{introduction}
 Quantum chromodynamics (QCD) is the
appropriate theory for describing the strong interaction at high
energy region, however, the strong gauge coupling at
low energy destroys the perturbative expansion  method.
 The long distance properties of QCD, especially the
 hadronic matrix elements can provide many important information for
 understanding and testing the standard model and beyond.
 The  exclusive semileptonic
decay $B\to \pi \bar{l} \nu_l$
can be used to determine the CKM parameter $|V_{ub}|$  \cite{Alexander}.
However, it requires a reliable
calculation of the form factor $f^+_{B\pi}(q^2)$
defined by
$ \langle\pi(p)\mid \bar{b}\gamma_\mu u\mid B(p+q)\rangle
=2f^+_{B\pi}(q^2) p_\mu +(f^+_{B\pi}(q^2)+f^-_{B\pi}(q^2)) q_\mu$,
with $p$ and $p+q$ being the $\pi$- and $B$-meson
four-momentum, respectively.
 $f^-_{B\pi}(q^2)$ plays a negligible role
for semileptonic decays into the light leptons $l=e,\mu$.

In Ref.\cite{Beneke}, the authors propose a formula called
 QCD factorization approach
for $B\rightarrow \pi \pi$, $\pi$K and $\pi$D to deal with
nonleptonic decays of B meson. In this
approach, the decay amplitudes are expressed in terms of the
semileptonic form factors, hadronic light-cone distribution
functions and hard-scattering  amplitudes.
The semileptonic form factors, the light-cone distribution  functions
are taken as input parameters and the hard-scattering amplitudes
are calculated by perturbative QCD.
 Again, the precise knowledge of heavy-to-light form factors
plays crucial roles. Among the existing approaches,
such as QCD sum rules, chiral perturbation theory ,
heavy quark effective theory  and phenomenological quark models,
the QCD light-cone sum rules (LCSR)  approach is very
 prominent for calculating $f^+_{B\pi}(q^2)$ \cite{Balitsky,Braun1,Chernyak1}.

The light-cone QCD sum rule approach carries out
 operator product expansion  near the light-cone $x^2\approx 0$
 instead of the short distance $x\approx 0$ while the nonperturbative
 matrices  are parameterized by light-cone wave functions
 which classified according to their twist instead of
 the vacuum condensates. For detailed discussion of this method ,
 one can see Ref. \cite{Kho1}. The LCSR for $f^+_{B \pi}(q^2)$ is valid at small and intermediate momentum transfer
squared
$ \label{chi}
q^2 \le m_Q^2 - 2 m_Q \chi ~, $
where $\chi$ is a typical hadronic scale of roughly $500 \; \mbox{MeV}$
and independent of the heavy quark mass $m_Q$.

In this paper, we calculate the form factor $f^+_{B\pi}(q^2)$
 (which is different from Refs. \cite{Kho2,Kho3,Kho4,Bagan}) up to twist-4
 light-cone functions by including
 perturbative $\alpha_s$-corrections for twist-2 terms
 using chiral current. Remarkably, the main uncertainties of the
 light-cone sum rules come from the light-cone wave functions.
   The chiral current approach has a striking advantage  that
 the twist-3 light-cone functions which are not known
 as well as the twist-2 light-cone functions eliminated and
 supposed to provide results with less uncertainties \cite{Huang}.
In fact, only the twist-2 wave function, which is dominant
in contributions to the sum rules, has
 been investigated systematically. The update investigation of
 the twist-3 wave functions can be found in Ref.\cite{Ball} and the
 calculations of the form factor $f^{+}_{B \pi}$
 including the $\alpha(s)$ corrections to the twist-3
 terms are performed in Ref.\cite{Ball2}.
 Although the QCD radiative corrections to the
twist-2 term for $f^{+}_{B \pi}$
are proven small in Ref. \cite{Kho3}, it is interesting to see whether
or not the case for chiral currents.

The article is organized as follows: correlator and sum
rule are derived in Sec.II; the perturbative
correlator are calculated to order $\alpha_s$ in Sec.III;
light-cone amplitudes and numerical results are presented in Sec.IV;
the section V is reserved for conclusion.

\section{correlator and sum rule }

Let us start with the following definition of $B\rightarrow \pi $ weak
form factors $f_{B \pi}(q^2)$:
\begin{equation}
\langle\pi (p)|\overline{u}\gamma _\mu b|B(p+q)\rangle=2f^{+}_{B \pi}(q^2)p_\mu
+(f^{+}_{B \pi} (q^2)+f^{-}_{B \pi} (q^2))q_\mu ,
\end{equation}
with $q$ being the momentum transfer. Following Ref. \cite{Huang},
we choose a chiral current to calculate the correlator function,
\begin{eqnarray}
\Pi_\mu (p,q) && =i\int d^4xe^{iqx}\langle\pi (p)|T
\{ \overline{u}(x)\gamma _\mu (1+\gamma
_5)b(x),\overline{b}(0)i(1+\gamma _5)d(0)\}|0\rangle m_b, \nonumber \\ &&
=\Pi (q^2,(p+q)^2)p_\mu +\widetilde{\Pi }(q^2,(p+q)^2)q_\mu ,
\end{eqnarray}
which is different from that in Ref. \cite{Kho2,Kho3,Kho4,Bagan}.
Here we take chiral limit $p^2=m_{\pi}^2=0$.

 We can insert  a complete series of intermediate states
with the same quantum numbers as the current operator
$\bar bi(1+\gamma _5)d$ in the correlator to obtain
the hadronic representation. After isolating the pole
term of the lowest pseudoscalar B meson, we get the result:
\begin{eqnarray}
\Pi_\mu ^H(p,q)  &&= \Pi ^H(q^2,(p+q)^2)p_\mu +
\widetilde{\Pi}^H(q^2,(p+q)^2)q_\mu,
 \nonumber \\
&&= \frac{ \langle \pi|\overline{u}\gamma _\mu b|B\rangle \langle
B|\overline{b}\gamma_5d|0\rangle m_b} {m_B^2-(p+q)^2} +\sum\limits_H\frac{\langle \pi |\overline{u}\gamma _\mu
(1+\gamma _5)|B_H\rangle \langle B_H|\overline{b}i(1+\gamma
_5)d|0\rangle m_b}{m_{B_H}^2-(p+q)^2}.
\end{eqnarray}

The intermediate states $B_H$ contain not only
pseudoscalar resonances  of the masses greater than $m_B$, but also
scalar resonances with $J^p=0^{+}$, corresponding to the operator
$\bar bd$. Taking into account the definition $\langle B|\bar{b}i\gamma
_5d|0\rangle={m_B}^2f_B/m_b$, we obtain:
\begin{eqnarray}
\Pi^H(q^2,(p+q)^2)&=&\frac{2 f^{+}_{B \pi}(q^2) m_B^2 f_B}{m_B^2-(p+q)^2}
+\int_{s_0}^{\infty}\frac
{\rho^H(s)}{s-(p+q)^2}ds, \nonumber \\
\widetilde{\Pi} ^H(q^2,(p+q)^2)&=&\frac{(f^{+}_{B \pi}(q^2)+f^{-}_{B \pi}(q^2)) m_B^2
f_B}{m_B^2 -(p+q)^2} +\int_ {s_0}^{\infty}\frac {\widetilde{\rho}
^H(s)}{s-(p+q)^2}ds.
\end{eqnarray}
Here  the contributions of higher resonances and continuum
states above the threshold $s_0$ are written in terms of
dispersion integrations, and the spectral densities
$\rho^H(s)$ and $\tilde{\rho}^H(s)$ can be
approximated by the quark-hadron duality ansatz. We can
 avoid the pollution from scalar resonances with $J^p=0^{+}$ by
 choosing $s_0$ near the B meson threshold and our final results
 conform this assumption.

In the following, we brief out line the calculation of the
correlator in QCD theory
and  work in the large space-like momentum
regions $(p+q)^2-m_b^2\ll 0$ for the $b\bar d$ channel, and $q^2\ll
m_b^2-O(1GeV^2)$ for the momentum transfer, which correspond to the small
light-cone distance $x^2\approx 0$ and are required by the validity of the
operator product expansion method. First,
we  write down the full $b$-quark propagator:

\begin{eqnarray}
\langle0|T{b(x)\bar{b}(0)}|0\rangle &&
=i\int \frac{d^4k}{(2\pi )^4}e^{-ikx}\frac{\hat
{k}+m}{k^2-m_b^2}-ig_s\int \frac{d^4k}{(2\pi
)^4}e^{-ikx}\int_0^1dv \left[\frac {1}{2}\frac{\hat{
k}+m}{(m_b^2-k^2)^2}G^{\mu \nu }(vx)\sigma _{\mu \nu }  \nonumber
\right. \\
&& \left. +\frac 1{m_b^2-k^2}vx_\mu G^{\mu \nu }(vx)\gamma _\nu
\right],
\end{eqnarray}
here $G_{\mu \nu }$ is the gluonic field strength, $g_s$ denotes the strong
coupling constant.
Substituting the above b quark propagator and the corresponding
$\pi$ meson light-cone wave functions into  Eq.(2) and
completing the integrations over $x$ and $k$, finally we obtain:
\begin{eqnarray}
\Pi(q^2,(p+q)^2) && =2f_\pi m_b^2 \int_{0}^{1} du \left
\{
\frac{\varphi _{\pi}(u)}{m_b^2-(1-u)q^2-u(p+q)^2}+
\frac{2ug_2(u)}{(m_b^2-(1-u)q^2-u(p+q)^2)^2} \nonumber  \right.\\
&&\left. -\frac{8m_b^2[g_1(u)+G_2(u)]}{(m_b^2-(1-u)q^2-u(p+q)^2)^3}
+ \int D\alpha_i \frac {2 \varphi_{\perp}(\alpha_i)
+2\tilde{\varphi}_{\perp}(\alpha_i)
-\varphi_{\parallel}(\alpha_i)-\tilde{\varphi}_{\parallel}(\alpha_i)}
{m_b^2-(1-\alpha_1-u\alpha_3)q^2-(\alpha_1+u\alpha_3)(p+q)^2} \right
\},
\end{eqnarray}
with $G_2(u)=-\int_{0}^{u} g_2(v) dv$ and $D\alpha_i=d\alpha_1 d\alpha_2 d\alpha_3 \delta(1-\alpha_1-\alpha_2
-\alpha_3)$. Here $\varphi_\pi$ is $\pi$ meson twist-2 light-cone
wave function, and $g_1(u),\ \ g_2(u),\ \ \varphi_{\perp}(\alpha_i),
\ \ \tilde{\varphi}_{\perp}(\alpha_i),\ \ \varphi_{\parallel}(\alpha_i),
\ \ \tilde{\varphi}_{\parallel}(\alpha_i)$ are $\pi$ meson
twist-4  light-cone wave functions. Their detailed expressions are given in
section 4.
Then we carry out the subtraction procedure of the continuum spectrum
by the standard procedure and perform the Borel transformations with
respect to $(p+q)^2$, and finally  obtain the result:
\begin{eqnarray}
f^{+}_{B\pi}(q^2) && = \frac{m_b^2 f_\pi}{m_B^2 f_B} e^{\frac{m_B^2}{M^2}}
\left\{\int_{\triangle}^{1} du e^{-\frac{m_b^2-q^2
(1-u)}{u M^2}} \left( \frac{\varphi_ \pi(u)}{u}
+\frac{2g_{2}(u)}{uM^2}-\frac{8m_b^2[g_1(u)+G_2(u)]}{u^3M^4}   \right) \nonumber \right. \\
&& \left.+\int_{0}^{1}
dv \int
D\alpha_i\frac {\theta(\alpha_1+v \alpha_3 -\Delta)}{(\alpha_1+v
\alpha_3)^2 M^2} e^{-\frac {m_b^2-(1-\alpha_1- v \alpha_3)
q^2}{M^2 (\alpha_1+v \alpha_3)}} \left( 2 \varphi_\perp(\alpha_i)+2
\widetilde\varphi_i\perp(\alpha_i)
-\varphi_\parallel(\alpha_i)-\widetilde\varphi_\parallel(\alpha_i)
\right )
 \right\}.
\end{eqnarray}
Here $\triangle=\frac{m_b^2-q^2}{s_0-q^2}$ and $s_0$ denotes the
subtraction of the continuum from the spectral integral. For
technical details, one can see Ref. \cite{Kho1,Huang}.

\section{Radiative corrections in order $\alpha_s$}

In this section, we calculate the perturbative contribution up to
 $\alpha_s$ for twist-2 terms, while the
 corrections for twist-3 terms and beyond are neglected, as they are
 supposed to be small.  Applying Borel
transformation for the $\alpha_s$ correction terms is tedious ,
we can  facilitate the calculation greatly by  writing
down the following dispersion integral relation:
\begin{eqnarray}
f^{+}_{B\pi}(q^2)=\frac{1}{2m^2_Bf_B}\int^{s_0}_{m_b^2} \rho^{QCD}(q^2,s)
e^{\frac{m^2_B-s}{M^2}}ds,
\end{eqnarray}
where
\begin{eqnarray}
\rho^{QCD}(q^2,s)
 = - \frac{f_{\pi}}{\pi}  \int_0^1 du \varphi_\pi (u)
\mbox{Im} T(q^2,s,u)~.
\end{eqnarray}
For example, with the zeroth order
approximation , one can easily obtain:
\begin{equation}
\mbox{Im} T_0(q^2,s,u)= -2\pi m_b^2 \delta(m_b^2-(1-u)q^2-us ).
\end{equation}
To  order $\alpha_s$, the amplitude can be written as
\begin{eqnarray}
T(r_1,r_2,u)=T_0(r_1,r_2,u)+\frac{\alpha_sC_F}{4\pi}T_1(r_1,r_2,u)~.
\end{eqnarray}
Here we introduce convenient dimensionless
variables $r_1 = q^2/m_b^2$
and $r_2=(p+q)^2/m_b^2$.
There are six  Feynman diagrams for
determining the first order amplitude
$T_1$ in perturbative expansion, .  For simplicity, we perform the calculation
in  Feynman gauge. In calculation, both the ultraviolet
  and collinear  divergences are regularized by
dimensional regularization and renormalized
in the $\overline{MS}$ scheme
with  totally anticommuting $\gamma_5$. To be more precise,
the collinear divergences in the hard amplitude are factored out
and absorbed in the evolution of the light-cone wave function
which is determined by the QCD evolution equation \cite{Brodsky}.
Finally we get the result:
\begin{eqnarray}
&T_1(r_1,r_2,u)&=
  \frac{6(1+\rho)}{(1-\rho)^2}
 \left(1- \ln\frac{m_b^2}{\mu^2} \right)
       -\frac{4}{1-\rho} \left[ ( G\left(\rho\right) -
 G\left(r_1\right))+( G\left(\rho\right)- G\left(r_2\right))
 \right]
\nonumber \\
 & &      +\frac{4}{(r_1-r_2)^2} \left(
             \frac{1-r_2}{u} \left[ G\left(\rho\right) -
G\left(r_1\right)\right]
           + \frac{1-r_1}{1-u} \left[ G\left(\rho\right) -
 G\left(r_2\right)\right] \right)  \nonumber\\
 & &      +2\frac{\rho+(1-\rho)\ln\left(1-\rho\right)}{\rho^2}
       -\frac{4}{1-\rho} \frac{(1-r_2)\ln\left(1-r_2\right)}{r_2}
+2\frac{3-\rho}{\left(1-\rho\right)^2}
\nonumber \\
 & &      -\frac{4}{(1-u)(r_1-r_2)}
             \left( \frac{ (1-\rho) \ln\left(1-\rho\right)}{\rho}
                  - \frac{ (1-r_2) \ln\left(1-r_2\right)}{r_2}
                  \right)
\end{eqnarray}
with
\begin{eqnarray}
&&\rho = r_1 + u (r_2-r_1), \ \ \mbox{Li}_2(x)=-\int^x_0\frac{dt}t \ln(1-t), \nonumber \\
G\left(\rho \right)
& = & \mbox{Li}_2(\rho) + \ln^2(1-\rho) -\ln(1-\rho)
\left(1 -\ln\frac{m_b^2}{\mu^2} \right).
\end{eqnarray}
As in the calculation of the non-leading order evolution kernel
of the wave function $\varphi_\pi(u,\mu)$, we take the
UV renormalization scale
and the factorization
scale of the collinear divergences to be  equal
\cite{Dittes,Sarmadi,Mikhailov}. Our results are of the same Dirac
structure as that of Ref.\cite{Kho3} but with different weight.

The $\overline{MS}$ quark mass depends explicitly on the
renormalization scale $\mu$ and implicitly on the renormalization
scheme. A renormalization scheme independent definition of the
quark mass within QCD perturbation theory
is given by the pole mass which is denoted by $m_b^*$.
As in Ref. \cite{Kho3}, we replace $\hat{m}_b$ by $m_b^*$
using the well-known one-loop relation:
\begin{eqnarray}
\hat{m}_b& = & m_b^*
\left \{ 1 + \frac{\alpha_S C_F}{4 \pi} \left(-4
         + 3 \ln \frac{m_b^{*2}}{\mu^2} \right) \right \}.
\end{eqnarray}
To $O(\alpha_s)$, this replacement adds a term,
\begin{eqnarray}
- \frac{4\rho}{(1-\rho)^2}  \left(
       4 - 3\ln\frac{m_b^{*2}}{\mu^2} \right),
\end{eqnarray}
to the renormalized amplitude
$T_1$.

To proceed further according to Eq.(9) we calculate
the imaginary part of the hard scattering amplitude
for $r_2>1$ and $r_1<1$:

$$-\frac{1}{\pi} \mbox{Im}T(r_1,r_2,u,\mu)=
 \frac{ \alpha_s(\mu) C_F }{2 \pi} \left\{
  \delta(1-\rho) \left[ 2\pi^2 -6 + 3\ln \frac{m_b^{*2}}{\mu^2}
 -2 \mbox{Li}_2(r_1) + 2 \mbox{Li}_2(\frac{1}{r_2})
 +\ln^2 r_2 \right.\right. $$
$$ \left. +2\frac{(1-r_2) \ln(r_2-1)}{r_2}
-2\ln^2(1-r_1)+2\ln(1-r_1)-2\ln(1-r_1)\ln \frac{m_b^{*2}}{\mu^2}
-2\ln^2(r_2-1)+2\ln(r_2-1)-2\ln(r_2-1)\ln \frac{m_b^{*2}}{\mu^2}
 \right] $$
$$+ \theta(\rho-1)\left[ 8 \left. \frac{\ln(\rho-1)}{\rho-1}
\right|_+
 +2 \left( \ln r_2+ \frac{1}{r_2} -2
-2 \ln(r_2-1) + \ln
\frac{m_b^{*2}}{\mu^2} \right)
             \left. \frac{1}{\rho-1} \right|_+
+ 2 \frac{1}{r_2-\rho}
 \left( \frac{1}{\rho} - \frac{1}{r_2} \right)
   +\frac{1-\rho}{\rho^2} \right.$$
$$ \left.
+2 \frac{1-r_1}{(r_1-r_2)(r_2-\rho)} \left( \ln \frac{\rho}
{r_2} -2 \ln \frac{\rho-1}{r_2-1} \right)
\left. -4 \frac{\ln \rho}{\rho-1}
-2 \frac{r_2-1}{(r_1-r_2)(\rho-r_1)}
   \left( \ln \rho -2 \ln(\rho-1) + 1 -
\ln \frac{m_b^{*2}}{\mu^2} \right )\right] \right. $$
$$+\theta(1-\rho) \left[ 2 \left( \ln r_2+ \frac{1}{r_2}
-2 \ln(r_2-1)
-\ln \frac{m_b^{*2}}{\mu^2}
\right) \left. \frac{1}{\rho-1} \right|_+
-2 \frac{1}{r_2-\rho}
\frac{1-r_2}{r_2} \right.$$
\begin{eqnarray}
\left. \left.
-2 \frac{1-r_1}{(r_1-r_2)(r_2-\rho)} \left( \ln r_2 + 1
 -2 \ln(r_2-1)
-\ln \frac{m_b^{*2}}{\mu^2}
\right) \right] \right\},
\end{eqnarray}
here, the operation $+$  is defined by
\begin{eqnarray}
F(x)|_+=\lim_{\eta\rightarrow 0} \left( F(x)\theta(1-x-\eta)-
\delta(1-x-\eta) \int_{0}^{1-\eta} F(y)dy \right),
\end{eqnarray}
and thus remove  the
spurious  divergences. The above expressions have a little difference
compared with the corresponding ones in Ref. \cite{Kho3} for coefficients
 of the $\delta(1-\rho)$ term.

Substituting Eq.(16) into Eq.(8-9), we can obtain
 the desired sum rule in $O(\alpha_s)$
for the form factor $f^+_{B\pi}$ in the leading twist-2
approximation.

\section{light-cone amplitudes and numerical results}

Let us choose  the input parameters entering the sum rule for
 $f^{+}_{B\pi}(q^2)$ first. To begin with, let us specify the
 pion wave functions.
For the leading twist-2 wave function $\varphi_\pi(u,\mu)$, the asymptotic
form is exactly
given by perturbative QCD  $\varphi_\pi(u,\mu \rightarrow
\infty)=6u(1-u)$ \cite{Brodsky,Chernyak2}, nonperturbative corrections can be included in a
systematic way in term of the approximate conformal invariance of QCD
and  expanded in terms of Gegenbauer
polynomials $C_{n}^{3/2}(2u-1)$ with weight $u(1-u)$. \\
To leading order (LO),
\begin{eqnarray}
\varphi_\pi(u,\mu)=6u(1-u) \sum\limits_{n} a_{n}(\mu_{0})
\left( \frac{\alpha_s(\mu)}{\alpha_s(\mu_0)}\right)^{\frac{\gamma^n_0}{\beta_0}} C_{n}^{3/2}
(2u-1);
\end{eqnarray}
and to non leading order (NLO)\cite{Kadantsera},
\begin{eqnarray}
\varphi_\pi(u,\mu)=6u(1-u) \sum\limits_{n} a_{n}(\mu_{0})
\exp\left(- \int_{\alpha_{s}(\mu_{0})}^{\alpha_{s}(\mu)} d\alpha
 \frac{\gamma^{n}(\alpha)}{\beta(\alpha)} \right)
 \left( C_{n}^{3/2}(2u-1) + \frac{\alpha_{s}(\mu)}{4 \pi}
 \sum\limits_{k>n} d_{n}^{k}(\mu) C_{k}^{3/2}(2u-1) \right),
\end{eqnarray}
with $a_0=1$.
Arguments based on conformal spin expansion
 allows one to neglect higher terms in this expansion and we take $n\leq4$.
The coefficients $a_2(\mu_0)=2/3$ and $a_4(\mu_0)=0.43$
at the scale $\mu_0=500$ MeV
have been extracted  from a two-point QCD sum rule
for the moments of $\varphi_\pi(u)$ \cite{Braun1,Chernyak2}.
The coefficients $d_n^k(\mu )$
are due to mixing effects, induced by the fact that
the polynomials $C_n^{3/2}(2u-1)$ weight by $u(1-u)$ are
the eigenfunctions of the LO, but not of
the NLO evolution kernel. The QCD
beta-function $\beta$
and the anomalous dimension $\gamma^n$ of the
$n$-th moment $a_n(\mu)$
of the wave function have to be taken in NLO \cite{Particle}.
We can substitute the corresponding values into the above
equation and obtain:
\begin{equation}
a_2(\mu_b)=0.35,\ \ a_4(\mu_b)=0.18 \ \ (LO); \ \ \ \ a_2(\mu_b)=0.218, \ \ a_4(\mu_b)=0.084 \ \ (NLO) ,
\end{equation}
at the scale $\mu_b=\sqrt{m_B^2-m_b^2} \approx 2.4 GeV$, which characterizes
the mean virtuality of the $b$ quark. The new analysis of the
experimental data on the $\gamma \gamma^* \pi$ and $\pi$
electromagnetic form factor indicates that the twist-2 wave function is
close to its asymptotic form \cite{Braun00}. In this article, we
use both nonasymptotic and asymptotic form for the $\pi$ twist-2
light-cone wave functions and compare the results.

The subleading twist-4 contributions are presently
known only in zeroth order in $\alpha_s$ \cite{Braun2,Gorsky}.
As the twist-3 contribution is eliminated, we
need only the twist-4 wave functions:
\begin{eqnarray}
&& \varphi_{\perp}(\alpha_i)=30 \delta^2 (\alpha_1-\alpha_2) \alpha_3^2
[\frac{1}{3}+2 \epsilon (1-2 \alpha_3)] , \ \ \
 \widetilde{\varphi}_{\perp}(\alpha_i)=30
\delta^2 \alpha_3^2 (1-\alpha_3) [\frac{1}{3}+2 \epsilon (1-2\alpha_3)] ,
\nonumber \\
&& \varphi_{\parallel}(\alpha_i)=120 \delta^2 \epsilon
(\alpha_1-\alpha_2)
\alpha_1 \alpha_2 \alpha_3 , \ \ \
 \tilde{\varphi}_{\parallel}(\alpha_i)=-120
\delta^2 \alpha_1 \alpha_2 \alpha_3 [\frac{1}{3}+\epsilon
(1-3\alpha_3)], \nonumber \\
&&g_1(u)  =
\frac{5}{2} \varepsilon^2 u^2 \overline{u}^2+\frac{1}{2} \varepsilon
\delta^2
 [u \overline{u} (2+13 u \overline{u})+10 u^3 \ln u (2-3
u+\frac{6}{5} u^2)
+10 \overline{u}^3\ln{\overline{u}}(2-3\overline{u}+\frac{6}{5}
 \overline{u}^2)] , \nonumber \\
 &&g_2(u)  =\frac{10}{3}\delta^2
u\overline{u}(u-\overline{u}).
\end{eqnarray}
with $\delta^2(\mu_b)=0.17GeV^2$ and $\varepsilon(\mu_b)=0.36$.
Unlike the case of the twist-2 wave functions, these twist-4 wave functions
seem to be very difficult to test by experiment, for they usually are of
negligible contributions in the sum rules.

Another important input is the decay constant of B meson $f_B$.
To keep consistently, we have to calculate  the two-point sum rule
for $f_B$ up to the corrections of order $\alpha_s$.
 Here we use the two-loop expression for the  running coupling constant
with $N_f=4$ and
$\bar{\Lambda}^{(4)}=234$ MeV
corresponding  to $\alpha_s(M_Z)= 0.112$ \cite{Particle} for comparing with
the results in Ref. \cite{Kho3}.
As the value of $\mu$  concerned , we take the value $2.4 GeV$ which
corresponding to the average virtuality
of the correlation function which  is given
by the Borel mass parameter $M^2$.
In the present case a chiral current correlator is adopted to
delete the contributions from the twist-3 wave functions,
we consider the following
two-point correlator:
\begin{equation}
\Pi(q^2)=i\int d^4xe^{iqx}\langle0|\overline{q}(x)(1+\gamma
_5)b(x),\overline{b}(0) (1-\gamma_5)q(0)|0\rangle.
\end{equation}
The standard manipulation yields three self-consistent sets of
results:
\begin{eqnarray}
f_B=218 MeV, & m_b=4.7GeV, & s_0=33GeV^2;  \nonumber \\
f_B=212MeV,&  m_b=4.8GeV, & s_0=34GeV^2;   \nonumber \\
 f_B=206MeV, & m_b=4.9GeV, & s_0=35GeV^2.
\end{eqnarray}
The corresponding $\alpha_s=0$ results:
\begin{eqnarray}
f_B=163 MeV, &  m_b=4.7GeV, & s_0=33GeV^2; \nonumber \\
 f_B=158MeV, & m_b=4.8GeV, & s_0=34GeV^2; \nonumber \\
 f_B=153MeV, & m_b=4.9GeV, & s_0=35GeV^2.
\end{eqnarray}
From the above results we can see that
$ \frac{f_B(\alpha_s=0)}{f_B(\alpha_s\neq 0)} \approx 76 \% $,
in other word, $\alpha_s$ corrections increase the value of $f_B$
about $30\%$.
They will be used as inputs in numerical analysis of
 the sum rule for $f^{+}_{B\pi}(q^2)$.  As for the B meson mass $m_B$
and the pion decay constant $f_\pi$, we
take the present world average value $m_B=5.279GeV$, and $f_\pi=0.132GeV$.
The continuum subtraction $s_0$ is about $33-35 GeV^2$ and the pole
mass for b quark is taken as $m_b=4.7-4.9GeV$. Here we make some
comments about the continuum subtraction $s_0$. The special chiral
current leads to cancellations between the condensates, the dominating
contributions come from the perturbative parts and
the nonperturbative parts only play tiny roles. The lowest
pseudoscalar resonance appears at the energy threshold about
$s=m_B^2 \approx 28 GeV^2$. Though the B meson has a narrow decay
width, the values taken in Ref. \cite{Huang} $s_0=30-33 GeV^2$ are
too low due to the large difference between the corresponding results for
the values of $f_B$.

We exploit
the sum rule numerically in the following: \\

\begin{eqnarray}
 f_Bf^{+}_{B\pi}(0)=60.5MeV,f^{+}_{B\pi}(0)=0.277, & m_b=4.7GeV, & s_0=33GeV^2;  \nonumber \\
 f_Bf^{+}_{B\pi}(0)=56.8MeV,f^{+}_{B\pi}(0)=0.268,&  m_b=4.8GeV, & s_0=34GeV^2;   \nonumber \\
 f_Bf^{+}_{B\pi}(0)=53.4MeV,f^{+}_{B\pi}(0)=0.259, & m_b=4.9GeV, &
 s_0=35GeV^2,
\end{eqnarray}
for $\alpha_s \neq 0$ in LO.

\begin{eqnarray}
 f_Bf^{+}_{B\pi}(0)=59.6MeV,f^{+}_{B\pi}(0)=0.273, & m_b=4.7GeV, &
 s_0=33GeV^2,
\end{eqnarray}
for  $\alpha_s \neq 0$ in NLO.

\begin{eqnarray}
 f_Bf^{+}_{B\pi}(0)=47.3MeV,f^{+}_{B\pi}(0)=0.290, &  m_b=4.7GeV, & s_0=33GeV^2; \nonumber \\
 f_Bf^{+}_{B\pi}(0)=44.1MeV,f^{+}_{B\pi}(0)=0.279, & m_b=4.8GeV, & s_0=34GeV^2; \nonumber \\
 f_Bf^{+}_{B\pi}(0)=41.2MeV,f^{+}_{B\pi}(0)=0.269, & m_b=4.9GeV, &
 s_0=35GeV^2,
\end{eqnarray}
for $\alpha_s=0$ in LO.
From the above results we can see that
$ \frac{f_Bf^{+}_{B\pi}(0)(\alpha_s\neq 0)}{f_Bf^{+}_{B\pi}(0)(\alpha_s = 0)} \approx 130 \% $,
in other word, $\alpha_s$ corrections increase the value of $f_Bf^{+}_{B\pi}(0)$
about $30\%$. Due to the same  corrections to the
decay constant, the resulting net $\alpha_s$ corrections are very
small, say, for $f^{+}_{B\pi}(0)$ less than $3 \%$.
The large correction for $f_B f^{+}_{B\pi}(0)$ is cancelled by
the corresponding value for $f_B$. They are compatible with
the values obtained in Ref. \cite{Kho3}, for $\alpha_s=0, f^+_{B\pi}(0)=0.30 $;
for $\alpha_s\neq0, f^+_{B\pi}(0)=0.27 $. Our numerical results show that
the vibrations for the form factor $f^+_{B\pi}(0)$ are about $\pm
0.01$ around the  center values, for
 $\alpha_s\neq0, f^+_{B\pi}(0)=0.27 $ ; for
$\alpha_s=0, f^+_{B\pi}(0)=0.28$ with LO wave functions.
It is shown in figure 1 that the form factor
$f^+_{B\pi}(q^2)$ with $\alpha_s$ corrections
 lies below the un-corrected one for LO  wave function (WF) ;
 the quantities of the $\alpha_s$ corrections increase with
$q^2$, at $q^2=15 GeV^2$, numerically lesser than $20 \%$ for LO wave functions;
 the curve for NLO wave function lies a little above the corresponding
 one for LO wave function; the curve for asymptotic wave function with
  $\alpha_s$ corrections almost the same as the un-corrected
  one for LO wave functions at $q^2 \succ 8 GeV^2$;
  the deviation of the curves for the $\alpha_s$ corrected LO wave function and asymptotic
  wave function from each other is notable.
In figure 2 and figure 3, we plot the $f^+_{B\pi}(q^2)$ as function
of $q^2$ in leading order $\pi$ light-cone wave function
with different boundary conditions. From two figures, we can see
that the vibrations for $f^+_{B\pi}(0)$ are small,  numerically
about $\pm0.01$ around the  center values
both for the $\alpha_s$ corrected and un-corrected
form factor.
 In figure 4, we use the parameters obtained
 in Ref.\cite{Braun00} as input, from the figure can see that
 the curve for  $f^+_{B\pi}(q^2)$ with $\alpha_s$ corrections varies
 according to the  $\pi$ twist-2 light-cone wave functions,
 the largest deviation of
 the values from each other is less than $15 \%$. In figure 5, we
 plot the $f^+_{B\pi}(q^2)$ with boundary condition $s_0=33GeV^2, m_b=4.7GeV$
 both for $\alpha_s$ corrected and un-corrected form factor using
the parameters obtained in Ref.\cite{Braun00}. Again, we can see
that the net $\alpha_s$ correction is small. There is a platform for
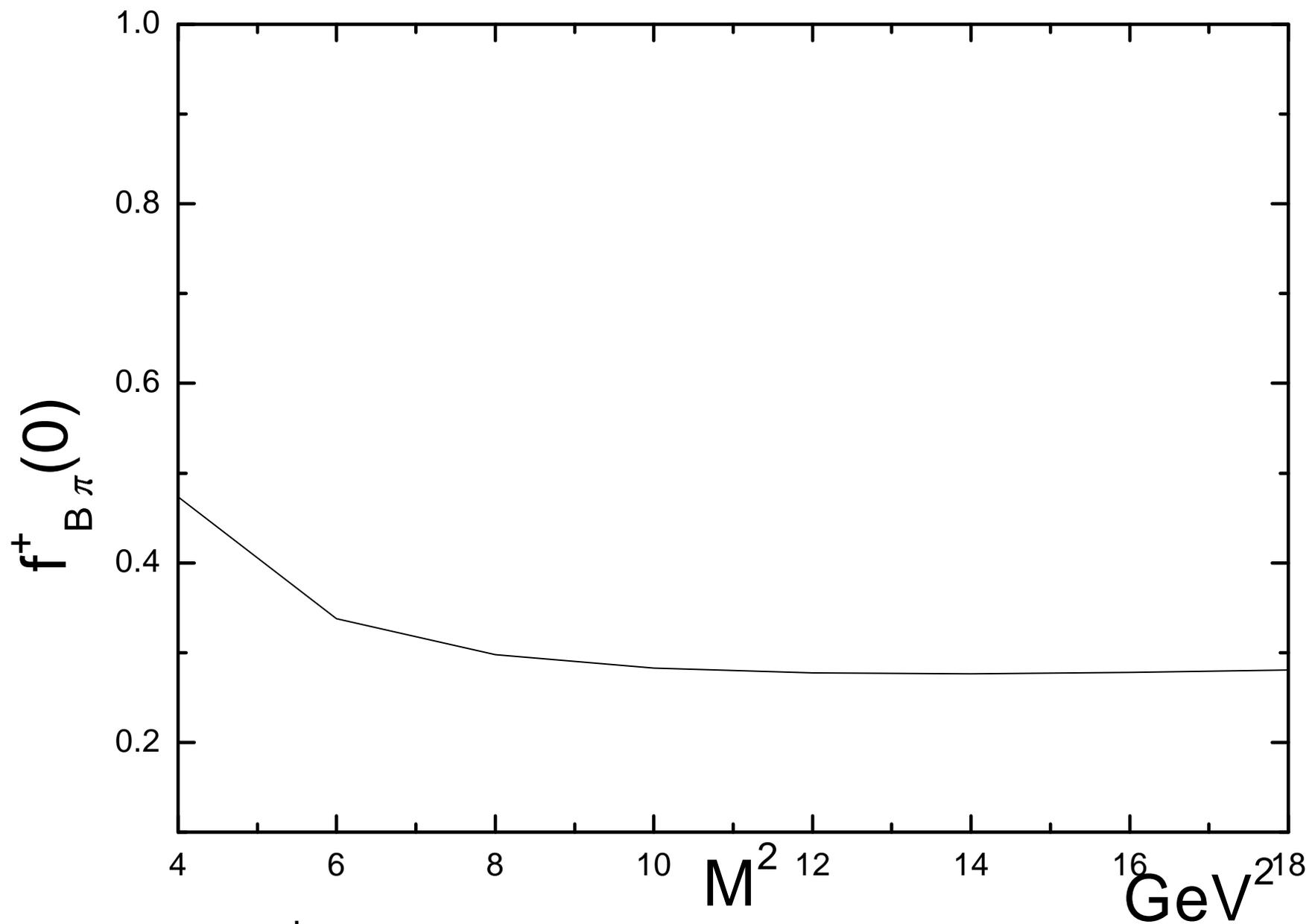
 $f^{+}_{B\pi}(q^2)$ as function of Borel parameter $M^2$ for
 $M^2= 8 - 14 GeV^2$ which verify
 the value we taken $M^2=12 GeV^2$ in calculation. For example,
the product $f^{+}_{B\pi}(0)$ is plotted as a
function of $M^2$ in figure 6. The uncertainties due to the
Borel parameter $M^2$ can thus be diminished or eliminated.

\section{ conclusion}
To summarize, we have re-examined that the weak form factor
$f^{+}_{B \pi}(q^2)$ up to $q^2=16GeV^2$ for B decays into light
pseudoscalar mesons by taking
the contributions of $\alpha_s$ corrections to twist-2 terms in
light-cone QCD sum rule framework. Due to the special structure of the
chiral current, the contributions of $\alpha_s$ corrections to twist-2
terms are of the same Dirac structure as that of Ref.\cite{Kho3}
with different weight.
As the contributions of twist-3 terms are eliminated, the
uncertainties due to the twist-3 light-cone wave functions
which are not understood as well as the twist-2
light-cone wave function  are avoided.
Furthermore, the possible pollution from wrong parity
$0^+$ mesons are deleted
by suitable choice of continuum subtraction parameter $s_0$,
the final results are
supposed to be with less uncertainties.
  The results presented here will be beneficial to
the precision extracting of the
CKM matrix element $|V_{ub}|$ from the exclusive processes $%
B\rightarrow\pi\ell\tilde{\nu_l}$ $(l=e,\mu)$, by confronting the
theoretical predictions with the experimentally available data.
Although the $\alpha_s$ corrections to $f_B f^+_{B\pi}$ are large,
about $30\%$, the similar corrections to $f_B$ canceled them, and
the resulting  net corrections to  form factor
$f^{+}_{B\pi}(q^2)$ are small. Our results are compatible with
the observations made in Ref.\cite{Kho3}.   Compared with the results obtained in
Ref.\cite{Kho3}, our results are with lesser uncertainties due to
the elimination of the twist-3 light-cone wave functions.
\section{acknowledgement}
This work is supported in part by National Science Foundation of
China.
One of the author (Z. G. Wang) would like to thank National
Postdoctoral Foundation for financial support.

\newpage

\end{document}